\documentclass[groupedaddress, superscriptaddress, citesort]{revtex4}
\usepackage{amssymb, amsmath}
\usepackage[dvips]{graphicx}

\def\be{\begin{equation}}
\def\ee{\end{equation}}
\def\ba{\begin{array}}
\def\ea{\end{array}}

\begin{document}
\baselineskip=18pt

\title {Dynamics of quantum coherence in Bell-diagonal states under Markovian channels}
\author{Yao-Kun Wang}
\affiliation{College of Mathematics,  Tonghua Normal University, Tonghua, Jilin 134001, China}
\affiliation{Research Center for Mathematics, College of Mathematics, Tonghua Normal University, Tonghua, Jilin 134001, China}
\author{Shao-Ming Fei}
\affiliation{School of Mathematical Sciences,  Capital Normal
University,  Beijing 100048,  China}
\affiliation{Max-Planck Institute for Mathematics in the Sciences, 04103 Leipzig, Germany}
\author{Zhi-Xi Wang}
\affiliation{School of Mathematical Sciences,  Capital Normal
University,  Beijing 100048,  China}

\begin{abstract}
We study the curves of coherence for the Bell-diagonal states including $l_{1}$-norm of coherence and
relative entropy of coherence under the Markovian channels in the first subsystem once. For a special Bell-diagonal state under bit-phase flip channel, we find frozen coherence under  $l_{1}$ norm occurs, but relative entropy of coherence decrease. It illustrates that the occurrence of frozen coherence depends on the type of the measure of coherence. We study the coherence evolution of Bell-diagonal states  under Markovian channels in the first subsystem $n$ times and find coherence under depolarizing channel decreases initially then increases for small $n$ and tend to zero for large $n$. We discuss the dynamics of coherence of the Bell-diagonal state under two independent same type local Markovian channels. We depict the dynamic behaviors of relative entropy of coherence for Bell-diagonal state under the bi-side different Markovian channel. We depict the dynamic behaviors of relative entropy of coherence for Bell-diagonal state under the bi-side different Markovian channel.
\end{abstract}

\maketitle

\section{Introduction}

Quantum coherence, stemming from quantum superposition rule, is thought of a special feature of quantum mechanic like entanglement and other quantum correlations. Quantum coherence is an essential ingredient in quantum information processing\cite{Bagan,Jha,Kammerlander}, quantum metrology\cite{Giovannetti,Demkowicz,Giovannetti1}, quantum optics\cite{Glauber,Sudarshan,Mandel}, low-temperature thermodynamics\cite{aberg,Narasimhachar,Oppenheim,Lostaglio,Lostaglio1,Vazquez,Wacker,pati}  and quantum biology\cite{Plenio,Rebentrost,Lloyd,Li,Huelga,levi}.
Recently, a framework to quantify coherence has been proposed\cite{Baumgratz}, and various quantum coherence measures, such as the $l_{1}$ norm of coherence\cite{Baumgratz}, relative entropy of coherence\cite{Baumgratz}, trace norm of coherence\cite{shao},  Tsallis relative $\alpha$ entropies\cite{Rastegin} and Relative R\'{e}nyi $\alpha$ monotones\cite{Chitambar}, have been defined.
By means of the coherence measures, a variety of properties of quantum coherence, such as the relations between quantum coherence and quantum correlations\cite{Ma,Radhakrishnan,Streltsov,Yao,Xi}, the freezing phenomenon of coherence\cite{Bromley,Yu}, have been investigated.

Quantum coherence is a useful physical resource. However, quantum systems inevitably subject to noise, which may result in the disappearance of coherence\cite{A. Streltsov-rev}. Similarly to the state protection in such a situation based on measurement reversal from weak measurement\cite{shuchaowang},
it is important to study the conditions under which the quantum coherence does not deteriorate under evolution of the system. Hence, the concept of frozen coherence was proposed, and the authors in Ref. \cite{Bromley} studied the dynamical conditions under which quantum coherence is totally
unaffected by quantum noise. For one qubit system, it has been shown that no nontrival condition exists such that the relative entropy coherence and the $l_1$ norm coherence are simultaneously frozen under any quantum channel. 
Experimentally, the froze phenomenon for relative entropy of coherence\cite{Baumgratz}, fidelity-based
measure of coherence\cite{Streltsov}, and trace norm
of coherence\cite{Bromley} for two and four qubits
under the phase damping channel were observed in a
nuclear magnetic resonance system\cite{silva}.

In this article, we investigate the coherence evolution in Bell-diagonal states based on $l_{1}$-norm of coherence and
relative entropy of coherence, when the first subsystem undergoes Markovian channels. We find that the relative entropy and the $l_{1}$ norm of coherence of the Bell-diagonal states under depolarizing channel decreases at first and then increases. In particular, we find the phenomenon of frozen coherence occurs for special Bell-diagonal states under bit-phase flip channel in terms of the $l_{1}$ norm of coherence.
It is shown that the occurrence of frozen coherence depends on the type of measures of coherence. We also study the coherence evolution of Bell-diagonal states when the first subsystem undergoes $n$ times Markovian channels. We find that the relative entropy of coherence for Bell-diagonal states under depolarizing channel decreases initially then increases for small $n$ and tends to $0$ for large $n$. We discuss the dynamics of coherence of the Bell-diagonal states under two independent local Markovian channels of the same type.

\section{Coherence evolution of Bell-diagonal states under Markovian channels with one parameter of decoherence probability $P$}

Under fixed reference basis, the $l_{1}$ norm of coherence of state $\rho$ is defined by
\begin{eqnarray}\label{l1}
C_{l_{1}}(\rho)=\sum_{i\neq j}|\rho_{i,j}|,
\end{eqnarray}
and the relative entropy of coherence is given by
\begin{eqnarray}\label{rel}
C_{r}(\rho)=S(\rho_{diag})-S(\rho),
\end{eqnarray}
where $S(\rho)=-Tr{\rho\log\rho}$ is von Neumann entropy.
Through out the paper, we take $\{|00\rangle$, $|01\rangle$, $|10\rangle$ and $|11\rangle\}$ the standard computational basis for two-qubit states.

A two-qubit Bell-diagonal state can be written as
\begin{eqnarray}\label{bs}
\rho=\frac{1}{4}(I\otimes I+\sum_{i=1}^3c_i\sigma_i\otimes\sigma_i),
\end{eqnarray}
where  $\{\sigma_i\}_{i=1}^3$ are the Pauli matrices.

\bigskip

(1) \emph{Coherence evolution of Bell-diagonal states under Markovian channels on the first subsystem}

\bigskip

We will consider the evolution of a quantum state
$\rho$ under a trace-preserving quantum operation $\varepsilon(\rho)$\cite{nielsen},
\begin{equation}\label{channel}
\varepsilon(\rho) = \sum_{i,j} \left(E_i\otimes E_j\right) \rho \left(E_i \otimes E_j\right)^\dagger,\nonumber
\end{equation}
where $\{E_k\}$ is the set of Kraus operators associated to a decohering process of a single qubit,
with $\sum_k E_k^\dagger E_k = I$. Typical channels are listed by the Kraus operators in Table~\ref{t1} \cite{wang}.

\begin{table}
\begin{center}
\begin{tabular}{c c}
\hline \hline
 & $\textrm{Kraus operators}$                                         \\  \hline & \\
BF   & $E_0 = \sqrt{1-p/2}\, I ,~~~ E_1 = \sqrt{p/2} \,\sigma_1$                        \\
 & \\
PF   & $E_0 = \sqrt{1-p/2}\, I ,~~~ E_1 = \sqrt{p/2}\, \sigma_3$                        \\
 & \\
BPF & $E_0 = \sqrt{1-p/2}\, I ,~~~ E_1 = \sqrt{p/2} \,\sigma_2$                        \\
 & \\
 DEP & $E_0 = \sqrt{1-p}\, I ,~~~ E_1 = \sqrt{p/3} \,\sigma_1$                        \\
 & \\
 & $E_2 = \sqrt{p/3}\,\sigma_2 ,~~~ E_3= \sqrt{p/3} \,\sigma_3$                        \\
 & \\
GAD   &
$E_0=\sqrt{p}\left(
\begin{array}{cc}
1 & 0 \\
0 & \sqrt{1-\gamma} \\
\end{array} \right) ,~~~~
E_2=\sqrt{1-p}\left(
\begin{array}{cc}
\sqrt{1-\gamma} & 0 \\
0 & 1 \\
\end{array} \right)$  \\
& \\
 & $E_1=\sqrt{p}\left(
\begin{array}{cc}
0 & \sqrt{\gamma} \\
0 & 0 \\
\end{array} \right) ,~~~~
E_3=\sqrt{1-p}\left(
\begin{array}{cc}
0 & 0 \\
\sqrt{\gamma} & 0 \\
\end{array} \right)$  \\ \hline \hline
\end{tabular}
\caption[table 1]{Kraus operators for the quantum channels: bit flip (BF), phase flip (PF),
bit-phase flip (BPF), depolarizing channel (DEP), and generalized amplitude damping (GAD), where
$p$ and $\gamma$ are decoherence probabilities, $0<p<1$, $0<\gamma<1$.}
\label{t1}
\end{center}
\end{table}

The decoherence processes BF, PF, BPF, and DEP in Table~\ref{t1} preserve the Bell-diagonal form of the
density operator $\rho$.
For the case of GAD, the Bell-diagonal form is kept for arbitrary $\gamma$
and $p=1/2$. In this situation, one has
\begin{equation}
\varepsilon(\rho)=\frac{1}{4}(I\otimes I+\sum_{i=1}^3c^\prime_i\sigma_i\otimes\sigma_i)\label{newstate},
\end{equation}
where the values of the $c^\prime_1$, $c^\prime_2$, $c^\prime_3$
are given in Table~\ref{t2} \cite{wang}. In the standard computational basis $\{|00\rangle, |01\rangle, |10\rangle, |11\rangle\}$, the density matrix has the form,
\begin{eqnarray}\label{newstate}
\rho^\prime= \frac{1}{4} \left(
\begin{array}{cccc}
1+c_3^\prime
& 0 & 0 & c_1^\prime -c_2^\prime \\
0 & 1-c_3^\prime & c_1^\prime+c_2^\prime & 0 \\
0 & c_1^\prime +c_2^\prime & 1-c_3^\prime
& 0 \\
c_1^\prime -c_2^\prime & 0 & 0 & 1+c_3^\prime
\end{array}
\right) \,,
\end{eqnarray}
where $c_{1}^\prime, c_{2}^\prime, c_{3}^\prime\in[-1,1]$.
From (\ref{l1}) and (\ref{rel}) we have
\begin{eqnarray}\label{newl1}
C_{l_{1}}(\rho^\prime)=\frac{1}{2}(|c_1^\prime +c_2^\prime|+|c_1^\prime -c_2^\prime|),
\end{eqnarray}
and
\begin{eqnarray}\label{rf}
C_{r}(\rho^\prime)&=&S(\rho^\prime_{diag})-S(\rho^\prime)\nonumber\\
&=&\frac{1}{4}(1-c_{1}^\prime-c_{2}^\prime-c_{3}^\prime)\log(1-c_{1}^\prime-c_{2}^\prime-c_{3}^\prime)\nonumber\\
 &+&\frac{1}{4}(1-c_{1}^\prime+c_{2}^\prime+c_{3}^\prime)\log(1-c_{1}^\prime+c_{2}^\prime+c_{3}^\prime)\nonumber\\
 &+&\frac{1}{4}(1+c_{1}^\prime-c_{2}^\prime+c_{3}^\prime)\log(1+c_{1}^\prime-c_{2}^\prime+c_{3}^\prime)\nonumber\\
 &+&\frac{1}{4}(1+c_{1}^\prime+c_{2}^\prime-c_{3}^\prime)\log(1+c_{1}^\prime+c_{2}^\prime-c_{3}^\prime)\nonumber\\
 &-&\frac{1+c_{3}^\prime}{2}\log(1+c_{3}^\prime)-\frac{1-c_{3}^\prime}{2}\log(1-c_{3}^\prime).
\end{eqnarray}

From Eq. (\ref{newl1}) and Eq. (\ref{rf}) we have the evolution of the $l_{1}$ norm coherence and the relative entropy of coherence for Bell-diagonal state, with $c_{1}=0.3$, $c_{2}=-0.4$ and $c_{3}=0.56$ under local nondissipative channels, see Fig. 1.

\begin{table}
\begin{center}
\begin{tabular}{c c c c}
\hline \hline
$\textrm{Channel}$ & $c^\prime_1$      & $c^\prime_2$     & $c^\prime_3$      \\ \hline
& & & \\
BF                 &  $c_1$            & $c_2 (1-p)^2$    & $c_3 (1-p)^2$     \\
& & & \\
PF                 &  $c_1 (1-p)^2$    & $c_2 (1-p)^2$    & $c_3$             \\
& & & \\
BPF                &  $c_1 (1-p)^2$    & $c_2$            & $c_3 (1-p)^2$     \\
& & & \\
DEP                &  $c_1 (1-4p/3)$   ~~~~~ & $c_2 (1-4p/3)$ ~~~~~       & $c_3 (1-4p/3)$     \\
& & & \\
GAD                &  $c_1 (1-p)$ & $c_2 (1-p)$ & $c_3 (1-p)^2$ \\ \hline \hline
\end{tabular}
\caption[table 2]{Correlation coefficients for the quantum operations: bit flip (BF), phase flip (PF),
bit-phase flip (BPF), depolarizing channel (DEP), and generalized amplitude damping (GAD). For GAD, we
have fixed $p=1/2$ and replaced $\gamma$ by $p$.}
\label{t2}
\end{center}
\end{table}

For amplitude damping channel the Kraus operators are given by,
\begin{eqnarray*}
E_0=\left(
\begin{array}{cc}
1& 0\\
0 & \sqrt{1-p}\\
\end{array}
\right),~~~~~
E_1=\left(
\begin{array}{cc}
0& \sqrt{p}\\
0 & 0\\
\end{array}
\right),
\end{eqnarray*}
$0 \leq p\leq 1$.
Under the amplitude damping channel the Bell-diagonal states are mapped to the output state $\rho_{adc}$,
\begin{eqnarray}\label{sta1}
\rho_{adc} = \frac{1}{4} \left(
\begin{array}{cccc}
2-(1-c_3)(1-p) & 0 & 0 & (c_1 -c_2)(1-p)^{\frac{1}{2}} \\
0 & 2-(1+c_3)(1-p) & (c_1 +c_2)(1-p)^{\frac{1}{2}} & 0 \\
0 & (c_1 +c_2)(1-p)^{\frac{1}{2}} & (1-c_3)(1-p) & 0 \\
(c_1 -c_2)(1-p)^{\frac{1}{2}} & 0 & 0 & (1+c_3)(1-p)
\end{array}
\right),
\end{eqnarray}
where $c_{1}, c_{2}, c_{3}\in[-1,1]$. The $l_{1}$ norm coherence and the relative entropy of coherence can be directly calculated by using Eq. (\ref{l1}) and Eq. (\ref{rel}) for $c_{1}=0.3$, $c_{2}=-0.4$ and $c_{3}=0.56$, see Fig. 1.

\begin{figure}[h]
\begin{center}
\scalebox{1.0}{(a)}{\includegraphics[width=7cm]{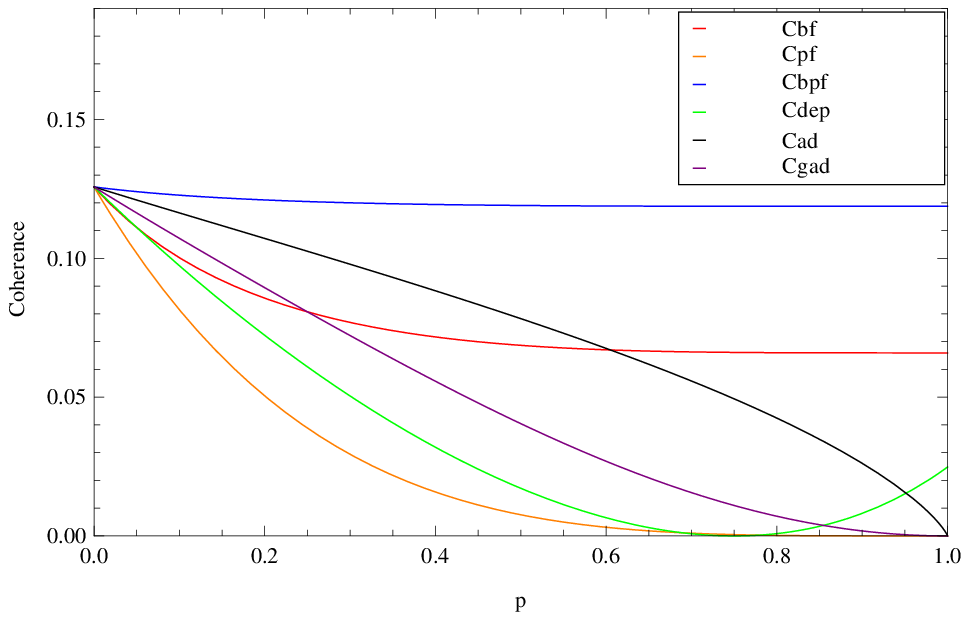}}
\scalebox{1.0}{(b)}{\includegraphics[width=7cm]{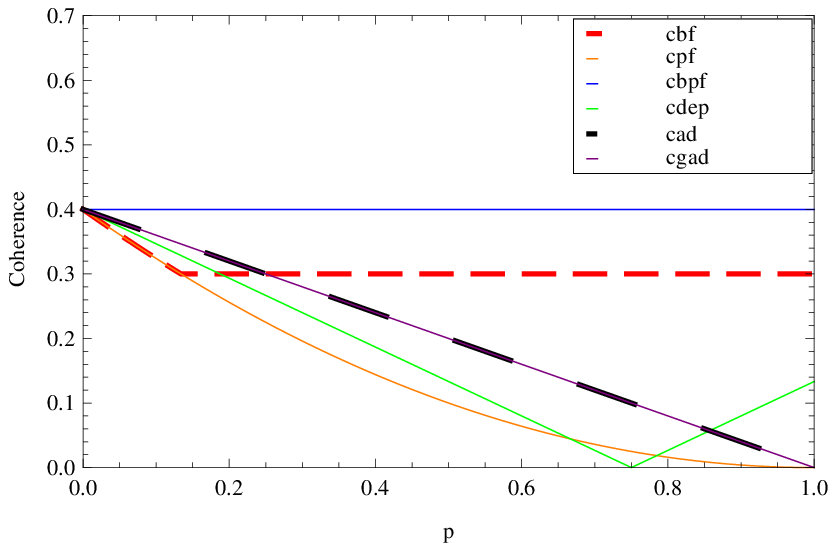}}
\caption{(a) Relative entropy of coherence for Bell-diagonal states \{$c_{1}=0.3, c_{2}=-0.4, c_{3}=0.56$\} under bit flip(Cbf), phase flip(Cpf), bit-phase flip(Cbpf), depolarizing(Cdep), amplitude damping(Cad), generalized amplitude damping(Cgad) as a function of $p$. (b) $l_{1}$ norm of coherence for Bell-diagonal states \{$c_{1}=0.3, c_{2}=-0.4, c_{3}=0.56$\} under bit flip(cbf), phase flip(cpf), bit-phase flip(cbpf), depolarizing(cdep), amplitude damping(cad), generalized amplitude damping(cgad).}
\end{center}
\end{figure}

We see that the coherence under the relative entropy and the $l_{1}$ norm of coherence behaviours similarly. Under the depolarizing channel, both the relative entropy and the $l_{1}$ norm of coherence decrease first and then increase. As $0\leq p\leq1$, according to Tabel \ref{t2} and Eq. (\ref{newl1}), the $l_{1}$ norm of coherence of the Bell-diagonal states under bit-phase flip channel is $\frac{1}{2}(|0.3(1-p)^{2}-0.4|+|0.3(1-p)^{2}+0.4|)=0.4$. Hence if the first subsystem goes through bit-phase flip channel, frozen coherence under  $l_{1}$ norm occurs[see blue line in Fig. 1(b)]. On the other hand, the relative entropy of coherence for the same Bell-diagonal state under bit-phase flip channel decreases as $p$ increases. Above all, the occurrence of frozen coherence depends on the type of measures of coherence. The same phenomenon shows up for the case of bit flip channel.

\bigskip

(2) \emph{Coherence evolution under $n$ times Markovian channels on the first subsystem}

\bigskip
Next, we consider the coherence dynamics of Bell-diagonal states under $n$ times Markovian channels on the first subsystem. As the decoherence processes BF, PF, BPF, DEP and GAD preserve the Bell-diagonal form of the density operator, if a Bell-diagonal state goes through the channel $n$ times, the parameters of the output state are given by $c^\prime_1$, $c^\prime_2$, $c^\prime_3$ given in tabel \ref{t3}. Using Eq. (\ref{rf}), we can obtain the relative entropy of coherence for Bell-diagonal state with \{$c_{1}=0.3$, $c_{2}=-0.4$ and $c_{3}=0.56$\} under $n$ times various Markovian noise channels, see Fig. 2.

\begin{table}
\begin{center}
\begin{tabular}{c c c c}
\hline \hline
$\textrm{Channel}$ & $c^\prime_1$      & $c^\prime_2$     & $c^\prime_3$      \\ \hline
& & & \\
$BF^{n}$                 &  $c_1$            & $c_2 (1-p)^{2n}$    & $c_3 (1-p)^{2n}$     \\
& & & \\
$PF^{n}$                 &  $c_1 (1-p)^{2n}$    & $c_2 (1-p)^{2n}$    & $c_3$             \\
& & & \\
$BPF^{n}$                &  $c_1 (1-p)^{2n}$    & $c_2$            & $c_3 (1-p)^{2n}$     \\
& & & \\
$DEP^{n}$  ~              &  $c_1 (1-4p/3)^{n}$ ~~~   & $c_2 (1-4p/3)^{n}$ ~~~       & $c_3 (1-4p/3)^{n}$     \\
& & & \\
$GAD^{n}$                &  $c_1 (1-p)^{n}$ & $c_2 (1-p)^{n}$ & $c_3 (1-p)^{2n}$ \\ \hline \hline
\end{tabular}
\caption[table 3]{Correlation coefficients for $n$ times quantum operations: bit flip ($BF^{n}$), phase flip ($PF^{n}$),
bit-phase flip ($BPF^{n}$), depolarizing channel ($DEP^{n}$), and generalized amplitude damping ($GAD^{n}$). For GAD, we have fixed $p=1/2$ and replaced $\gamma$ by $p$.}
\label{t3}
\end{center}
\end{table}

On the other hand, if the first subsystem of Bell-diagonal state goes through the amplitude damping channel, the output state $\rho_{adc}^{(1)}$ is given by
\begin{eqnarray}
\rho_{adc}^{(1)}=E_0\otimes I \rho^{ab} E_0^\dagger\otimes I + E_1\otimes I \rho^{ab} E_1^\dagger\otimes I,
\end{eqnarray}
If the first subsystem goes through this channel twice, the output state $\rho_{adc}^{(2)}$ is as follows:
\begin{eqnarray}
\rho_{adc}^{(2)}=E_0\otimes I \rho_{adc}^{(1)} E_0^\dagger\otimes I + E_1\otimes I \rho_{adc}^{(1)} E_1^\dagger\otimes I.
\end{eqnarray}
If the first subsystem goes through the amplitude damping channel $n$ times, then the output state $\rho_{adc}^{(n)}$ is of the form
\begin{eqnarray}
\rho_{adc}^{(n)}=E_0\otimes I \rho_{adc}^{(n-1)} E_0^\dagger\otimes I + E_1\otimes I \rho_{adc}^{(n-1)} E_1^\dagger\otimes I,
\end{eqnarray}
which can be rewritten as
\begin{eqnarray}
\rho_{adc}^{(n)}=\sum_{i_1,i_2,\cdots, i_n=0,1}  E_{i_1i_2\cdots i_n} \otimes I \rho^{ab} E_{i_1i_2\cdots i_n}^\dagger\otimes I
\end{eqnarray}
with $E_{i_1i_2\cdots i_n}=E_{i_1}E_{i_2}\cdots E_{i_n}$. Due to the properties of operators $E_0$ and $E_1$ in the amplitude damping channel,
\begin{eqnarray}
E_1^2=0, \ \ E_0E_1=E_1,\ \ E_1E_0=\sqrt{1-p}E_1,
\end{eqnarray}
$\rho_{adc}^{(n)}$ is reduced to the following form,
\begin{eqnarray}
\rho_{adc}^{(n)}=E_0^n \otimes I \rho^{ab} (E_0^n)^\dagger\otimes I+\sum_{i=0}^{n-1} E_1 E_0^{n-i-1} \otimes I \rho^{ab} (E_1 E_0^{n-i-1})^\dagger \otimes I.
\end{eqnarray}

By straightforward calculation, we obtain
\begin{eqnarray}\label{left rho}
\rho_{adc}^{(n)}=\frac{1}{4}\left(
\begin{array}{cccc}
2-(1-c_{3})(1-p)^{n}& 0 & 0 & (c_{1}-c_{2})(1-p)^{\frac{n}{2}}\\
0& 2-(1+c_{3})(1-p)^{n} & (c_{1}+c_{2})(1-p)^{\frac{n}{2}}& 0\\
0& (c_{1}+c_{2})(1-p)^{\frac{n}{2}} & (1-c_{3})(1-p)^{n} & 0\\
(c_{1}-c_{2})(1-p)^{\frac{n}{2}}& 0 & 0 & (1+c_{3})(1-p)^{n}
\end{array}
\right).
\end{eqnarray}

As an example, the relative entropy of coherence of $\rho_{adc}^{(n)}$ for Bell-diagonal state \{$c_{1}=0.3$, $c_{2}=-0.4$ and $c_{3}=0.56$\} is shown in Fig. 2 (f).
One can see that when $p$ approaches to $1$, the relative entropy of coherence approaches to a constant for large $n$, that is, frozen coherence almost appears, see Fig. 2 (a). A similar behaviour can be seen in the relative entropy of coherence for Bell-diagonal states under $n$ times bit-phase flip channel. But these coherences gather together independent of $n$, see Fig. 2 (c). When $p$ increases, the relative entropy of coherence for Bell-diagonal state under depolarizing channel decreases initially and then increases. It is worth mentioning that the coherence tends to $0$ as $p$ approaches to $1$ for large $n$, see Fig. 2 (d). Relative entropy of coherence for Bell-diagonal states under $n$ times phase flip channel and generalized amplitude damping channel decreases as $p$ increases. The curvature gradually become large for large $n$, see Fig. 2 (b) and (e). When $p$ increases, the relative entropy of coherence for Bell-diagonal states under $n$ times amplitude damping channel also decreases. But the curvature gets smaller at first and then larger as $n$ increases, see Fig. 2 (f).

\begin{figure}[h]
\begin{center}
\scalebox{1.0}{(a)}{\includegraphics[width=4.8cm]{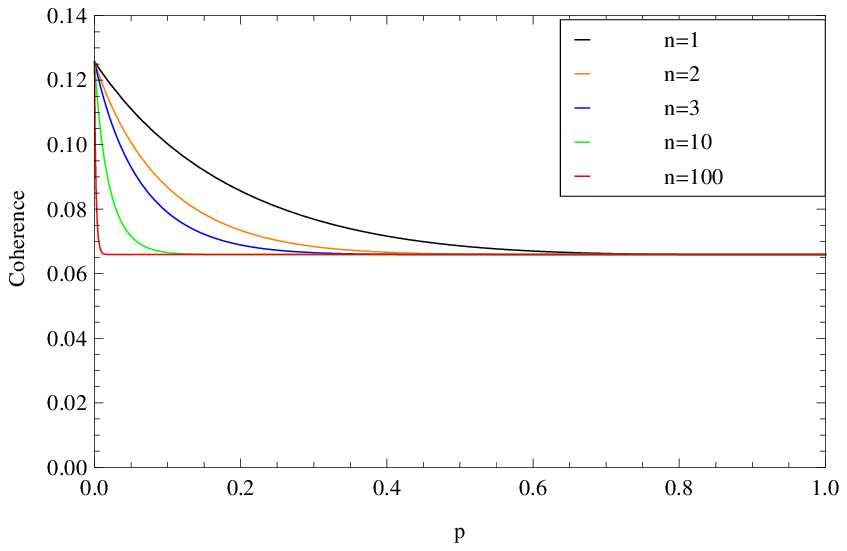}}
\scalebox{1.0}{(b)}{\includegraphics[width=4.8cm]{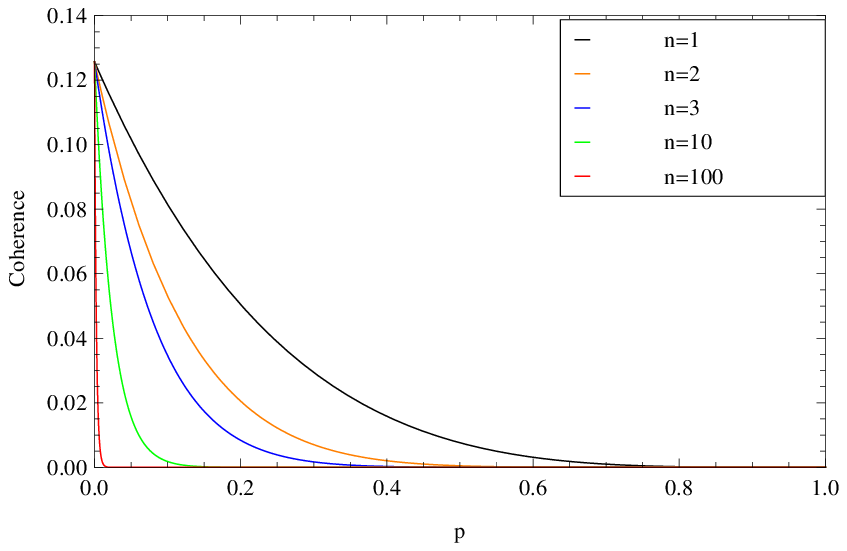}}
\scalebox{1.0}{(c)}{\includegraphics[width=4.8cm]{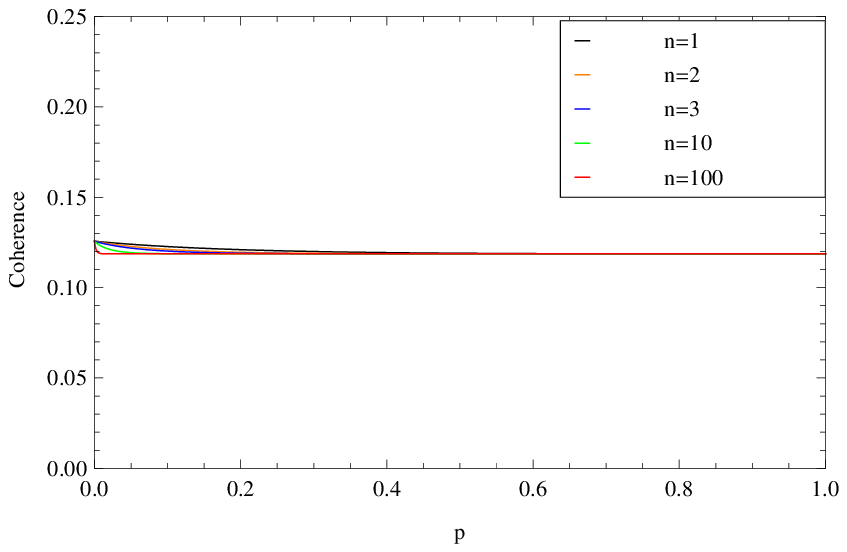}}
\scalebox{1.0}{(d)}{\includegraphics[width=4.8cm]{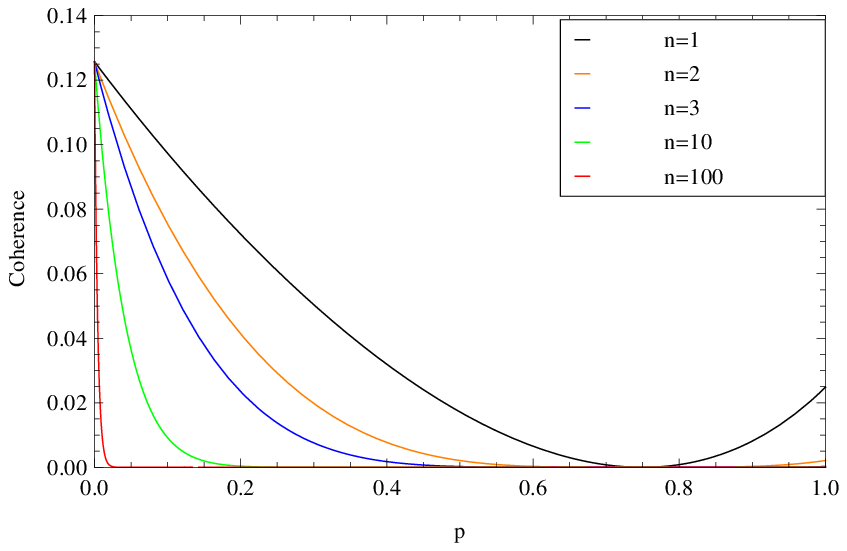}}
\scalebox{1.0}{(e)}{\includegraphics[width=4.8cm]{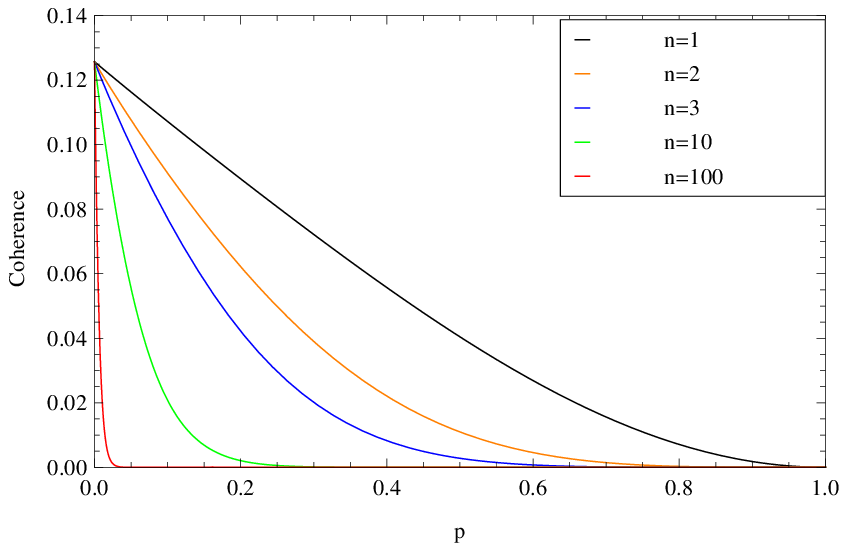}}
\scalebox{1.0}{(f)}{\includegraphics[width=4.8cm]{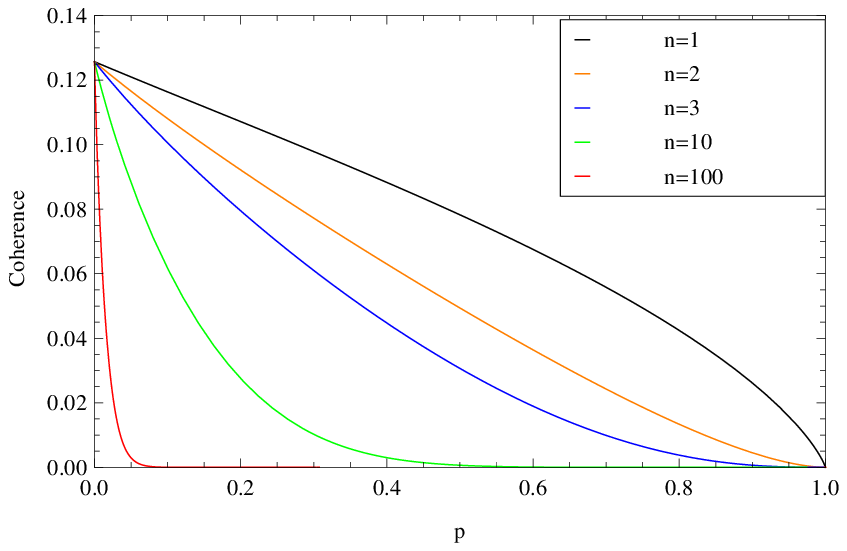}}
\caption{Relative entropy of coherence for Bell-diagonal state \{$c_{1}=0.3$, $c_{2}=-0.4$ and $c_{3}=0.56$\}: (a) under bit flip channel $n$ times, (b) under  phase flip channel $n$ times, (c) under bit-phase flip channel $n$ times, (d) under depolarizing channel $n$ times, (e) under generalized amplitude damping channel $n$ times, (f) under amplitude damping channel $n$ times.}
\end{center}
\end{figure}

\bigskip

(3) \emph{Coherence evolution under $n$ times amplitude damping channel on both two subsystems of Bell-diagonal states}

\bigskip

Furthermore, if two subsystems both go through the amplitude damping channel $n$ times, the output state is given by
\begin{eqnarray}
\rho^{(n)}=\sum_{i_1,i_2,\cdots, i_n,j_1,j_2,\cdots, j_n=0,1} E_{i_1i_2\cdots i_n}\otimes E_{j_1j_2\cdots j_n}  \rho E_{i_1i_2\cdots i_n}^\dagger\otimes E_{j_1j_2\cdots j_n}^\dagger.
\end{eqnarray}
By straightforward calculation, we obtain the output state $\rho_{adc}^{(n,n)}$ of Bell-diagonal state  under bi-side amplitude damping channel,

\begin{eqnarray}\label{left rho}
\rho_{adc}^{(n,n)}=\frac{1}{4}\left(
\begin{array}{cccc}
x& 0 & 0 & (c_{1}-c_{2})(1-p)^{n}\\
0& y & (c_{1}+c_{2})(1-p)^{n}& 0\\
0& (c_{1}+c_{2})(1-p)^{n} & y & 0\\
(c_{1}-c_{2})(1-p)^{n}& 0 & 0 & (1+c_{3})(1-p)^{2n}
\end{array}
\right),
\end{eqnarray}
where $x=4-4(1-p)^{n}+(1+c_{3})(1-p)^{2n}$, $y=2(1-p)^{n}-(1+c_{3})(1-p)^{2n}$.

Using Eq. (\ref{rel}), we have the relative entropy of coherence for the case that both of the subsystems of Bell-diagonal states undergo $n$ times amplitude damping channel, see Fig. 3, for \{$c_{1}=0.3$, $c_{2}=-0.4$ and $c_{3}=0.56$\}. We find that the relative entropy of coherence decreases as $p$ increases. As $n$ increases, the curves have larger curvature.

\begin{figure}[h]
\begin{center}
\scalebox{1.0}{\includegraphics[width=9cm]{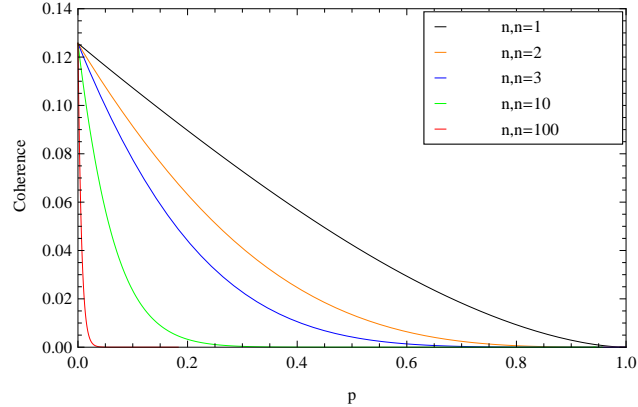}}
\caption{Relative entropy of coherence for that both subsystems of Bell-diagonal states \{$c_{1}=0.3, c_{2}=-0.4, c_{3}=0.56$\} undergo the amplitude damping channel $n$ times: n=1(black line), n=2(orange line), n=3(blue line), n=10(green line), n=100(red line).}
\end{center}
\end{figure}

\section{Coherence dynamics of Bell-diagonal states under Bi-side Markovian channels of the same type}

In this section, we discuss the dynamics of coherence of a
two-qubit Bell-diagonal state undergoing two independent local Markovian channels of the
same type but of different decoherence rates. There are three types of
local Markovian channels: bit-flip noise, phase-flip
noise, and bit-phase-flip, described by the Kraus operators in Table~\ref{t4}, respectively.

\begin{table}
\begin{center}
\begin{tabular}{c c}
\hline \hline
 & $\textrm{Kraus operators}$                                         \\  \hline & \\
PF-PF   & $E_0^{(A)} = \sqrt{1-\frac{p}{2}} I^{A}\otimes I^{B},~~~ E_1^{(A)}=\sqrt{\frac{p}{2}}\sigma^A_3\otimes I^{B}$             \\
 & \\
  & $E_0^{(B)} =I^{A}\otimes\sqrt{1-\frac{q}{2}}I^{B},~~~  E_1^{(B)} =I^{A}\otimes\sqrt{\frac{q}{2}}\sigma^B_3$                 \\
 & \\

BF-BF   & $E_0^{(A)}=\sqrt{1-\frac{p}{2}}I^{A}\otimes I^{B},~~~ E_1^{(A)}=\sqrt{\frac{p}{2}}\sigma^A_1\otimes I^{B}$                \\
 & \\
 & $E_0^{(B)}=I^{A}\otimes\sqrt{1-\frac{q}{2}}I^{B},~~~ E_1^{(B)}=I^{A}\otimes\sqrt{\frac{q}{2}}\sigma^B_1$                 \\
 & \\

 BPF-BPF   & $E_0^{(A)}=\sqrt{1-\frac{p}{2}}I^{A}\otimes I^{B},~~~ E_1^{(A)}=\sqrt{\frac{p}{2}}\sigma^A_2\otimes I^{B}$                \\
 & \\
 & $E_0^{(B)}=I^{A}\otimes\sqrt{1-\frac{q}{2}}I^{B},~~~ E_1^{(B)}=I^{A}\otimes\sqrt{\frac{q}{2}}\sigma^B_2$                 \\
 & \\

 \hline \hline
\end{tabular}
\caption[table 1]{Kraus operators for two independent local Markovian channels: two independent local phase-flip channels (PF-PF), two independent local bit-flip channels(BF-BF), two independent local bit-phase-flip channels (BPF-BPF), where $p=1-\text{exp}(-\gamma t)$, $q=1-\text{exp}(-\gamma't)$, and
$\gamma$ and $\gamma'$ are the phase damping rates for the channels
on the qubits $A$ and $B$, respectively.}
\label{t4}
\end{center}
\end{table}

Any Bell-diagonal state $\rho$ given by (\ref{bs}) evolves to another Bell-diagonal state Eq. (\ref{newstate}) under these channels. The corresponding coefficients are listed on the  Table \ref{t5}. As an example, from Eq. (\ref{rf}) we have the dynamical behaviors of the relative entropy of coherence for Bell-diagonal state \{$c_{1}=0.3$, $c_{2}=-0.4$ and $c_{3}=0.56$\} under the bi-side same type Markovian channel of bit flip (cbf), phase flip (cpf), bit-phase flip (cbpf), see Fig. 4.
Interestingly, the corresponding coherences have the relation, cbfp$>$cbf$>$cpf. When $p$ and $q$ increase, the coherence decrease and the coherence under phase flip channel tends to $0$.

\begin{table}
\begin{center}
\begin{tabular}{c c c c}
\hline \hline
$\textrm{Channel}$ & $c^\prime_1$      & $c^\prime_2$     & $c^\prime_3$      \\ \hline
& & & \\
PF-PF                 &  $(1-p)(1-q)c_1$~~~~            & $(1-p)(1-q)c_2$    & $c_3$     \\
& & & \\
BF-BF                 &  $c_1$    &~~~ $(1-p)(1-q)c_2$ ~~~   & $(1-p)(1-q)c_3$             \\
& & & \\
BPF-BPF        ~~~        &  $(1-p)(1-q)c_1$    & $c_2$            & $(1-p)(1-q)c_3$     \\
& & & \\
 \hline \hline
\end{tabular}
\caption[table 2]{Correlation coefficients for the quantum operations: two independent local phase-flip channels (PF-PF), two independent local bit-flip channels(BF-BF), two independent local bit-phase-flip channels (BPF-BPF), where $p=1-\text{exp}(-\gamma t)$, $q=1-\text{exp}(-\gamma't)$, and
$\gamma$ and $\gamma'$ are the phase damping rates for the channels
on the qubits $A$ and $B$, respectively.}
\label{t5}
\end{center}
\end{table}

\begin{figure}[h]
\begin{center}
\scalebox{1.0}{\includegraphics[width=7cm]{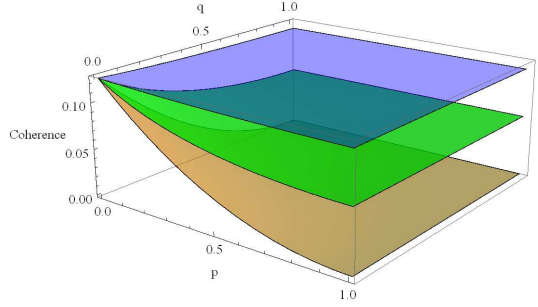}}
\caption{Relative entropy of coherence for Bell-diagonal state \{$c_{1}=0.3, c_{2}=-0.4, c_{3}=0.56$\} under bi-side Markovian channels of the same type: bit flip channel (cbf) [green surface], phase flip channel (cpf) [orange surface], bit-phase flip channel (cbpf) [blue surface]}.
\end{center}
\end{figure}

Furthermore, if both two subsystems go through the same type channel $n$ times, the Bell-diagonal state $\rho$ given by (\ref{bs}) also evolves to another Bell-diagonal state (\ref{newstate}). The corresponding coefficients are listed in Table \ref{t51}. As an example, from Eq. (\ref{rf}) we obtain the relative entropy of coherence for Bell-diagonal state \{$c_{1}=0.3, c_{2}=-0.4, c_{3}=0.56$\} under bi-side same type Markovian channel $n$ times. In Fig. 5, we can see that when $n$ becomes larger, the relative entropy of coherence for Bell-diagonal state under bi-side same type Markovian channel tends to a constant quickly as $p$ and $n$ increase.

\begin{table}
\begin{center}
\begin{tabular}{c c c c}
\hline \hline
$\textrm{Channel}$ & $c^\prime_1$      & $c^\prime_2$     & $c^\prime_3$      \\ \hline
& & & \\
$PF^{n}-PF^{n}$                 &  $(1-p)^{n}(1-q)^{n}c_1$            & $(1-p)^{n}(1-q)^{n}c_2$    & $c_3$     \\
& & & \\
$BF^{n}-BF^{n}$                &  $c_1$    & $(1-p)^{n}(1-q)^{n}c_2$    & $(1-p)^{n}(1-q)^{n}c_3$             \\
& & & \\
$BPF^{n}-BPF^{n}$                &  $(1-p)^{n}(1-q)^{n}c_1$    & $c_2$            & $(1-p)^{n}(1-q)^{n}c_3$     \\
& & & \\
 \hline \hline
\end{tabular}
\caption[table 2]{Correlation coefficients for $n$ times quantum operations: two independent local phase-flip channels ($PF^{n}-PF^{n}$), two independent local bit-flip channels($BF^{n}-BF^{n}$), two independent local bit-phase-flip channels ($BPF^{n}-BPF^{n}$), where $p=1-\text{exp}(-\gamma t)$, $q=1-\text{exp}(-\gamma't)$, and $\gamma$ and $\gamma'$ are the phase damping rates for the channels
on qubits $A$ and $B$, respectively.}
\label{t51}
\end{center}
\end{table}

\begin{figure}[h]
\begin{center}
\scalebox{1.0}{(a)}{\includegraphics[width=6cm]{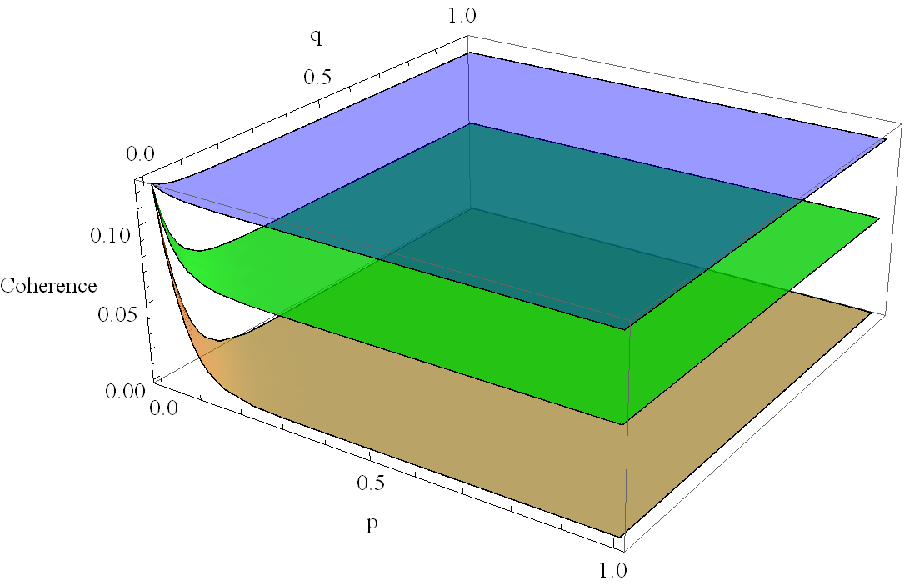}}
\scalebox{1.0}{(b)}{\includegraphics[width=6cm]{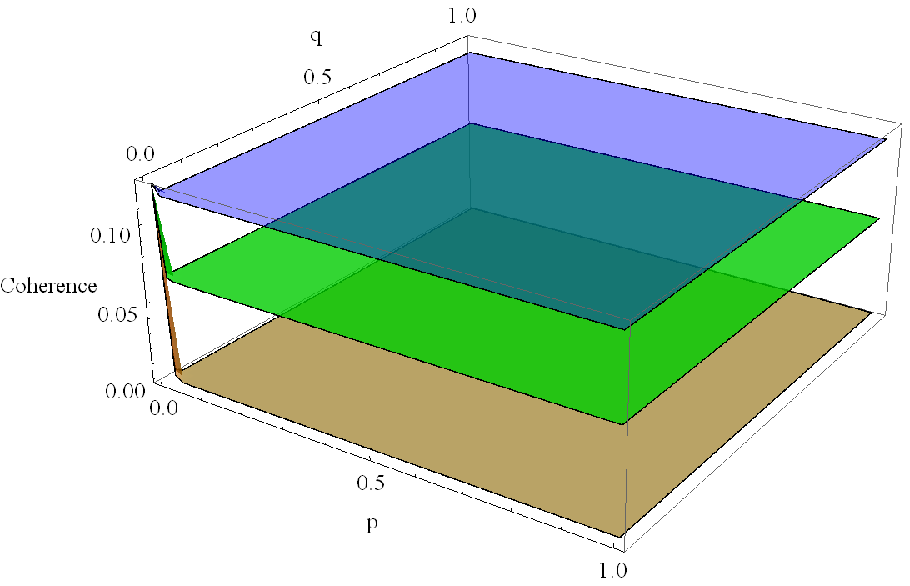}}
\caption{Relative entropy of coherence for Bell-diagonal state \{$c_{1}=0.3, c_{2}=-0.4, c_{3}=0.56$\} under $n$ times bi-side Markovian channel of the same type: bit flip channel (cbf) [green surface], phase flip channel (cpf) [orange surface], bit-phase flip channel (cbpf) [blue surface], (a) $n=10$ and (b) $n=100$.}
\end{center}
\end{figure}

\section{Coherence dynamics of Bell-diagonal states under bi-side Markovian channels of different types}

In this section, we discuss the dynamics of coherence of a
two-qubit Bell-diagonal state under two different local
Markovian channels. From the phase-flip channel, bit-flip channel and bit-phase-flip
channel given by the Kraus operators in Table~\ref{t6}, respectively, any Bell-diagonal state $\rho$ given by (\ref{bs}) evolves to another Bell-diagonal state (\ref{newstate}) under two different local Markovian channels. The corresponding coefficients are listed in Table \ref{t7}. As an example, from (\ref{rf}) the dynamical behavior of the relative entropy of coherence for Bell-diagonal state \{$c_{1}=0.3, c_{2}=-0.4, c_{3}=0.56$\} under the bi-side different Markovian channels are shown in Fig. 6. It can be seen that when $p$ and $q$ increase, the relative entropy of coherence decreases under all bi-side channels. In particular, when $q$ and $p$ approach to $1$, the coherences for Cbf-pf and Cpf-bpf tend to $0$, see Fig. 6 (a) and (c). When $q$ and $p$ approach $1$ simultaneously , the coherences for Cbf-bpf approaches $0$, see Fig. 6 (b).

\begin{table}
\begin{center}
\begin{tabular}{c c}
\hline \hline
 & $\textrm{Kraus operators}$                                         \\  \hline & \\
BF-PF   & $E_0^{(A)}=\sqrt{1-\frac{p}{2}}I^{A}\otimes I^{B},~~~~ E_1^{(A)}=\sqrt{\frac{p}{2}}\sigma^A_1\otimes I^{B},$             \\
 & \\
  & $E_0^{(B)}=I^{A}\otimes\sqrt{1-\frac{q}{2}}I^{B},~~~~  E_1^{(B)}=I^{A}\otimes\sqrt{\frac{q}{2}}\sigma^B_3$                 \\
 & \\
BF-BPF   & $E_0^{(A)}=\sqrt{1-\frac{p}{2}}I^{A}\otimes I^{B},~~~~ E_1^{(A)}=\sqrt{\frac{p}{2}}\sigma^A_1\otimes I^{B},$                \\
 & \\
 & $E_0^{(B)}=I^{A}\otimes\sqrt{1-\frac{q}{2}}I^{B},~~~~ E_1^{(B)}=I^{A}\otimes\sqrt{\frac{q}{2}}\sigma^B_2$                 \\
 & \\
PF-BPF   & $E_0^{(A)}=\sqrt{1-\frac{p}{2}}I^{A}\otimes I^{B},~~~~ E_1^{(A)}=\sqrt{\frac{p}{2}}\sigma^A_3\otimes I^{B},$                \\
 & \\
 & $E_0^{(B)}=I^{A}\otimes\sqrt{1-\frac{q}{2}}I^{B},~~~~ E_1^{(B)}=I^{A}\otimes\sqrt{\frac{q}{2}}\sigma^B_2$                 \\
 & \\

 \hline \hline
\end{tabular}
\caption[table 1]{Kraus operators for two-qubit sysems under two different local Markovian channels: a bit-flip channel and a phase-flip channel (BF-PF), a bit-flip channel and a bit-phase-flip channel(BF-BPF), a phase-flip channel and a bit-phase-flip channel (PF-BPF), where $p=1-\text{exp}(-\gamma t)$, $q=1-\text{exp}(-\gamma't)$, and $\gamma$ and $\gamma'$ are the phase damping rates for the channels
on qubits $A$ and $B$, respectively.}
\label{t6}
\end{center}
\end{table}

\begin{table}
\begin{center}
\begin{tabular}{c c c c}
\hline \hline
$\textrm{Channel}$ & $c^\prime_1$      & $c^\prime_2$     & $c^\prime_3$      \\ \hline
& & & \\
BF-PF                 &  $(1-q)c_1$            & $(1-p)(1-q)c_2$    & $(1-p)c_3$     \\
& & & \\
BF-BPF                 &  $(1-q)c_1$    & $(1-p)c_2$    &~~~ $(1-p)(1-q)c_3$~~~             \\
& & & \\
PF-BPF~~~                &  $(1-p)(1-q)c_1$    & $(1-p)c_2$            & $(1-q)c_3$     \\
& & & \\
 \hline \hline
\end{tabular}
\caption[table 2]{Correlation coefficients for the quantum operations: a bit-flip channel and a phase-flip channel (BF-PF), a bit-flip channel and a bit-phase-flip channel(BF-BPF), a phase-flip channel and a bit-phase-flip channel (PF-BPF), where $p=1-\text{exp}(-\gamma t)$, $q=1-\text{exp}(-\gamma't)$, and
$\gamma$ and $\gamma'$ are the phase damping rates for the channels on qubits $A$ and $B$, respectively.}
\label{t7}
\end{center}
\end{table}

\begin{figure}[h]
\begin{center}
\scalebox{1.0}{(a)}{\includegraphics[width=4.8cm]{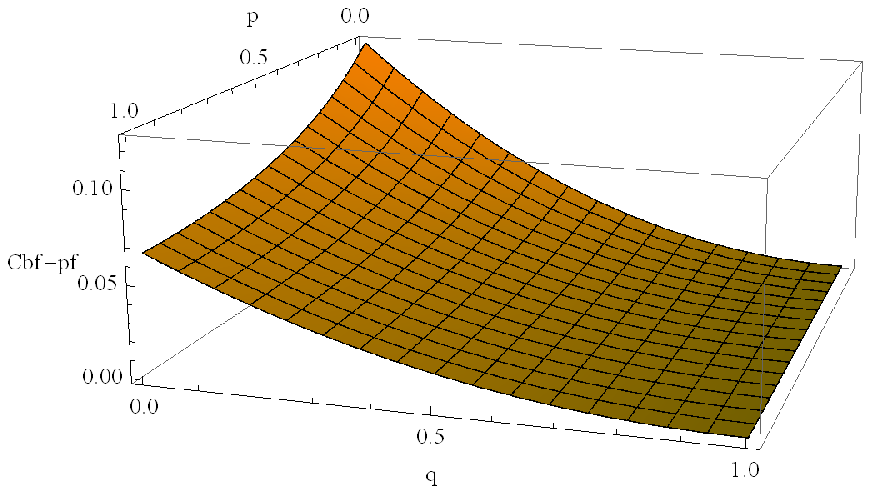}}
\scalebox{1.0}{(b)}{\includegraphics[width=4.8cm]{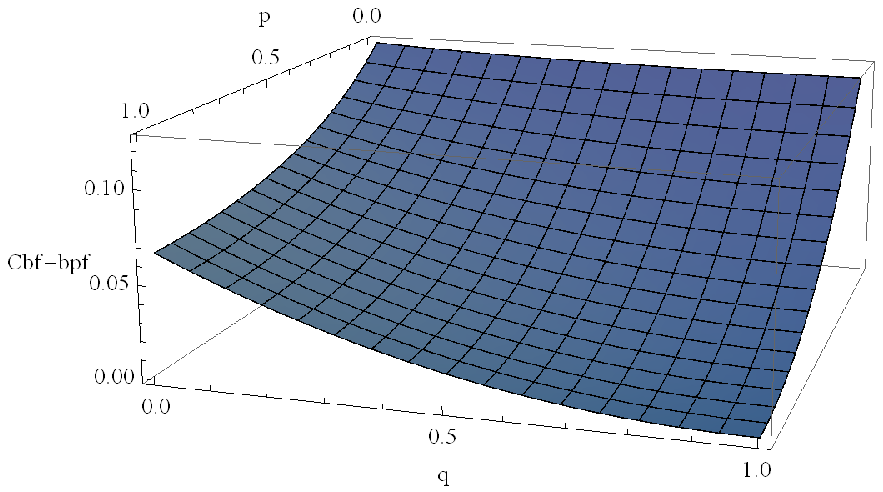}}
\scalebox{1.0}{(c)}{\includegraphics[width=4.8cm]{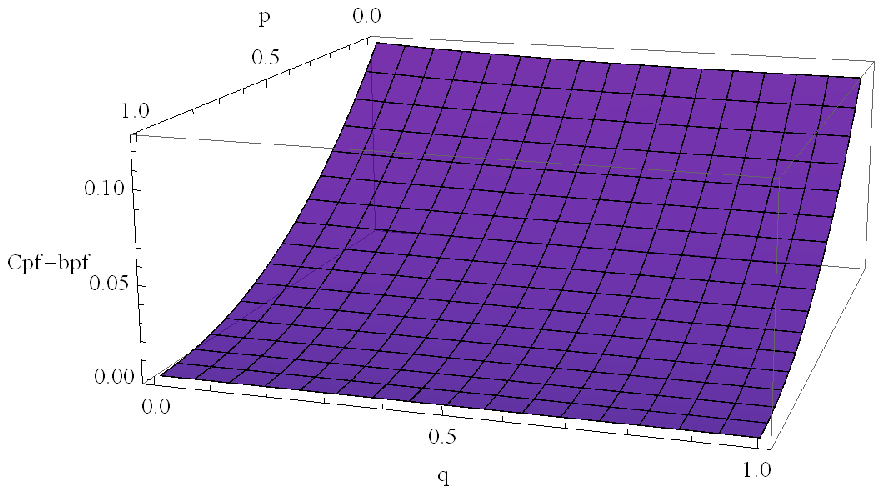}}
\caption{For Bell-diagonal state \{$c_{1}=0.3, c_{2}=-0.4, c_{3}=0.56$\}: (a) Relative entropy of coherence  under the bit-flip channel and  phase-flip channel as a function of $p$ and $q$ (Cbf-pf). (b) Relative entropy of coherence under the bit-flip channel and bit-phase-flip channel as a function of $p$ and $q$ (Cbf-bpf). (c) Relative entropy of coherence under the phase-flip channel and bit-phase-flip channel as a function of $p$ and $q$ (Cpf-bpf).}
\end{center}
\end{figure}

Furthermore, we discuss the dynamics of coherence of a two-qubit Bell-diagonal state under two different local Markovian channels $n$ times. Any Bell-diagonal state $\rho$ given by (\ref{bs}) evolves to another Bell-diagonal state (\ref{newstate}) under  two different local Markovian channels $n$ times. The corresponding coefficients are listed in Table \ref{t8}. In Fig. 7, as an example, from Eq. (\ref{rf}) we have the dynamical behaviors of the relative entropy of coherence for Bell-diagonal state \{$c_{1}=0.3, c_{2}=-0.4, c_{3}=0.56$\} under the bi-side different Markovian channels $10$ times. Comparing with Fig. 6, as soon as $p$ and $q$ increase, the relative entropy of coherence under all the bi-side channel tends to $0$ suddenly. The changing trend of the relative entropy of coherence is controlled by $q$, see Fig. 7 (a). The relative entropy of coherence under the phase-flip channel and bit-phase-flip channel $n=10$ times is controlled by $p$, see Fig. 7 (c). Soon as $q$ and $p$ increase simultaneously, the relative entropy of coherence under the bit-flip channel and bit-phase-flip channel $n=10$ times approaches $0$, see Fig. 7 (b).

\begin{table}
\begin{center}
\begin{tabular}{c c c c}
\hline \hline
$\textrm{Channel}$ & $c^\prime_1$      & $c^\prime_2$     & $c^\prime_3$      \\ \hline
& & & \\
$BF^{n}-PF^{n}$                 &  $(1-q)^{n}c_1$            & $(1-p)^{n}(1-q)^{n}c_2$    & $(1-p)^{n}c_3$     \\
& & & \\
$BF^{n}-BPF^{n}$                 &  $(1-q)^{n}c_1$    & $(1-p)^{n}c_2$    & $(1-p)^{n}(1-q)^{n}c_3$             \\
& & & \\
$PF^{n}-BPF^{n}$                &  $(1-p)^{n}(1-q)^{n}c_1$    & $(1-p)^{n}c_2$            & $(1-q)^{n}c_3$     \\
& & & \\
 \hline \hline
\end{tabular}
\caption[table 2]{Correlation coefficients for the $n$ times quantum operations: a bit-flip channel and a phase-flip channel ($BF^{n}-PF^{n}$), a bit-flip channel and a bit-phase-flip channel($BF^{n}-BPF^{n}$), a phase-flip channel and a bit-phase-flip channel ($PF^{n}-BPF^{n}$), where $p=1-\text{exp}(-\gamma t)$, $q=1-\text{exp}(-\gamma't)$, and
$\gamma$ and $\gamma'$ are the phase damping rates for the channels on qubits $A$ and $B$, respectively.}
\label{t8}
\end{center}
\end{table}

\begin{figure}[h]
\begin{center}
\scalebox{1.0}{(a)}{\includegraphics[width=4.8cm]{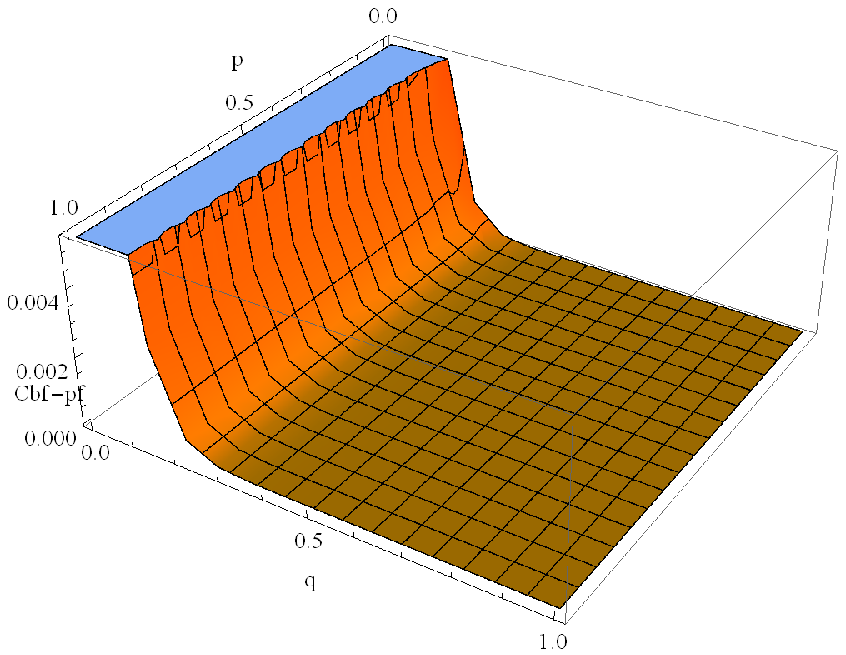}}
\scalebox{1.0}{(b)}{\includegraphics[width=4.8cm]{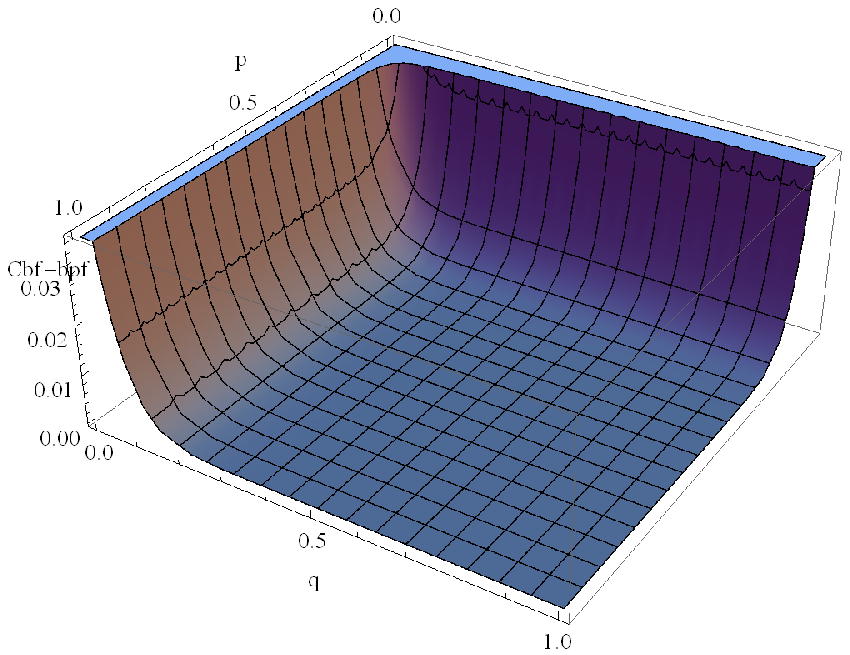}}
\scalebox{1.0}{(c)}{\includegraphics[width=4.8cm]{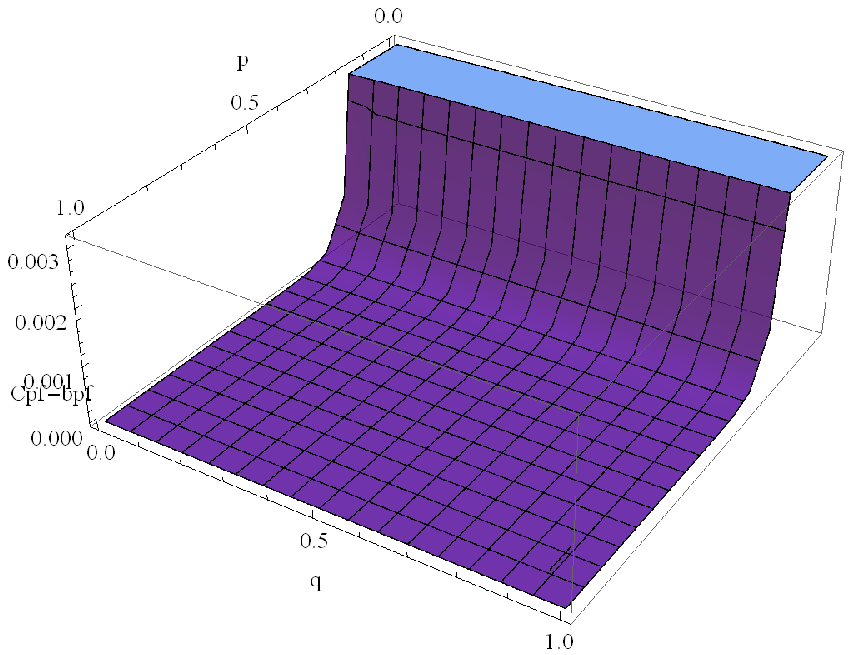}}
\caption{For Bell-diagonal state \{$c_{1}=0.3, c_{2}=-0.4, c_{3}=0.56$\}: (a) Relative entropy of coherence  under the bit-flip channel and  phase-flip channel $n=10$ times (Cbf-pf). (b) Relative entropy of coherence under the bit-flip channel and bit-phase-flip channel $n=10$ times (Cbf-bpf) (c) Relative entropy of coherence under the phase-flip channel and bit-phase-flip channel $n=10$ times (Cpf-bpf).}
\end{center}
\end{figure}

\section{\bf summary}\label{IIII}

In this work, we have investigated the $l_{1}$-norm of coherence and the
relative entropy of coherence under the channels of bit flip, phase flip, bit-phase flip, depolarizing, amplitude damping, generalized amplitude damping on the first subsystem for Bell-diagonal states.
In particular, we find that the $l_{1}$ norm of coherence of the special Bell-diagonal state under bit-phase flip channel is constant: frozen coherence under $l_{1}$ norm occurs. On the other hand, the relative entropy of coherence for the same Bell-diagonal state under bit-phase flip channel decreases as $p$ increases. It has been shown that the occurrence of frozen coherence depends on the type of measures of coherence.

We have studied the coherence evolution under Markovian channels on the first subsystem $n$ times.
We have discussed the dynamics of coherence of the Bell-diagonal states under two independent same type local Markovian channels of bit-flip, phase-flip, and bit-phase-flip.
We have shown the dynamical behaviors of the relative entropy of coherence for Bell-diagonal states under the bi-side different Markovian channel.
Furthermore, we have discussed the dynamics of coherence of Bell-diagonal states under two different local Markovian channels $n$ times.
Our results highlight the investigations of coherence evolution under local quantum channels.

\bigskip
\noindent {\bf Acknowledgments}  This work was supported by the National Natural Science Foundation of China under grant No. 11675113, and NSF of Beijing under No. KZ201810028042.


\begin{thebibliography}{18}




\bibitem{Bagan} E. Bagan, J. A. Bergou, S. S. Cottrell, and M. Hillery, Phys. Rev. Lett. \textbf{116}, 160406 (2016).
\bibitem{Jha} P. K. Jha, M. Mrejen, J. Kim, C. Wu, Y. Wang, Y. V. Rostovtsev, and X. Zhang, Phys. Rev. Lett. \textbf{116}, 165502 (2016).
\bibitem{Kammerlander} P. Kammerlander and J. Anders, Sci. Rep. \textbf{6}, 22174 (2016).
\bibitem{Giovannetti} V. Giovannetti, S. Lloyd, and L. Maccone, Science \textbf{306}, 1330 (2004).
\bibitem{Demkowicz} R. Demkowicz-Dobrza{\'n}ski and L. Maccone, Phys. Rev. Lett. \textbf{113}, 250801 (2014).
\bibitem{Giovannetti1} V. Giovannetti, S. Lloyd, and L. Maccone, Nat. Photonics \textbf{5}, 222 (2011).
\bibitem{Glauber} R. J. Glauber, Phys. Rev. \textbf{131}, 2766 (1963).
\bibitem{Sudarshan} E. C. G. Sudarshan, Phys. Rev. Lett. \textbf{10}, 277 (1963).
\bibitem{Mandel} L. Mandel and E. Wolf, \emph{Optical Coherence and Quantum Optics} (Cambridge University Press, Cambridge, UK, 1995).
\bibitem{aberg} J. Aberg, Phys. Rev. Lett. \textbf{113}, 150402 (2014).
\bibitem{Narasimhachar} V. Narasimhachar and G. Gour, Nat. Commun. \textbf{6}, 7689 (2015).
\bibitem{Oppenheim} P. {\'C}wikli{\'n}ski, M. Studzi{\'n}ski, M. Horodecki, and J. Oppenheim, Phys. Rev. Lett. \textbf{115}, 210403 (2015).
\bibitem{Lostaglio} M. Lostaglio, D. Jennings, and T. Rudolph, Nat. Commun. \textbf{6}, 6383 (2015).
\bibitem{Lostaglio1} M. Lostaglio, K. Korzekwa, D. Jennings, and T. Rudolph, Phys. Rev. X \textbf{5}, 021001 (2015).
\bibitem{Vazquez}H. Vazquez, R. Skouta, S. Schneebeli, M. Kamenetska, R. Breslow, L. Venkataraman, and M. S. Hybertsen, Nat. Nanotechnol. \textbf{7}, 663 (2012).
\bibitem{Wacker} O. Karlstr\"{o}m, H. Linke, G. Karlstr\"{o}m, and A. Wacker, Phys. Rev. B \textbf{84}, 113415 (2011).
\bibitem{pati} A. Misra, U. Singh, S. Bhattacharya, and A. K. Pati, Phys. Rev. A \textbf{93}, 052335 (2016).
\bibitem{Plenio} M. B. Plenio and S. F. Huelga, New J. Phys. \textbf{10}, 113019 (2008).
\bibitem{Rebentrost} P. Rebentrost, M. Mohseni, and A. Aspuru-Guzik, J. Phys. Chem. B \textbf{113}, 9942 (2009).
\bibitem{Lloyd} S. Lloyd, J. Phys: Conf. Ser. \textbf{302}, 012037 (2011).
\bibitem{Li} C.-M. Li, N. Lambert, Y.-N. Chen, G.-Y. Chen, and F. Nori, Sci. Rep. \textbf{2}, 885 (2012).
\bibitem{Huelga} S. F. Huelga and M. B. Plenio, Contemp. Phys. \textbf{54}, 181 (2013).
\bibitem{levi} F. Levi, and F. Mintert, New J. Phys. \textbf{16}, 033007 (2014).
\bibitem{Baumgratz} T. Baumgratz, M. Cramer, and M. B. Plenio, Phys. Rev. Lett. \textbf{113}, 140401 (2014).
\bibitem{shao} L.-H. Shao, Z. Xi, H. Fan and Y. Li, Phys. Rev. A \textbf{91} 042120 (2015).
\bibitem{Rastegin} A. E. Rastegin, Phys. Rev. A \textbf{93} 032136 (2016).
\bibitem{Chitambar} E. Chitambar and  G. Gour, Phys. Rev. A \textbf{94} 052336 (2016).
\bibitem{Ma} J. Ma, B. Yadin, D. Girolami, V. Vedral, and M. Gu, Phys. Rev. Lett. \textbf{116}, 160407 (2016).
\bibitem{Radhakrishnan} C. Radhakrishnan, M. Parthasarathy, S. Jambulingam, and T. Byrnes, Phys. Rev. Lett. \textbf{116}, 150504 (2016).
\bibitem{Streltsov} A. Streltsov, U. Singh, H. S. Dhar, M. N. Bera, and G. Adesso, Phys. Rev. Lett. \textbf{115}, 020403 (2015).
\bibitem{Yao} Y. Yao, X. Xiao, L. Ge, and C. P. Sun, Phys. Rev. A \textbf{92}, 022112 (2015).
\bibitem{Xi} Z. Xi, Y. Li, and H. Fan, Sci. Rep. \textbf{5}, 10922 (2015).
\bibitem{Bromley} T. R. Bromley, M. Cianciaruso, and G. Adesso, Phys. Rev. Lett. \textbf{114}, 210401 (2015).
\bibitem{Yu} X.-D. Yu, D.-J. Zhang, C. L. Liu, and D. M. Tong, Phys. Rev. A \textbf{93}, 060303 (2016).
\bibitem{A. Streltsov-rev} A. Streltsov, G. Adesso, M. B. Plenio, Rev. Mod. Phys. \textbf{89}, 041003.
\bibitem{shuchaowang} S.-C. Wang, Z.-W. Yu, W.-J. Zhou, X.-B. Wang, Phys. Rev. A \textbf{89}, 022318 (2014).
\bibitem{silva} I. A. Silva, A. M. Souza, T. R. Bromley, M. Cianciaruso, R. Marx,
R. S. Sarthour, I. S. Oliveira, R. L. Franco, S. J. Glaser, E. R.
deAzevedo, D. O. Soares-Pinto, G. Adesso, Phys. Rev. Lett. \textbf{117}, 160402 (2016).
\bibitem{nielsen} M. A. Nielsen, and I. L. Chuang,  \emph{Quantum Computation and Quantum Information} (Cambridge University Press,  Cambridge,  UK,  2000).
\bibitem{wang} Y.-k. Wang, S.-M. Fei, Z.-X. wang, J.-P. Cao, H. Fan, Sci. Rep. \textbf{10}, 10727 (2015).




\end{thebibliography}
\end{document}